\documentclass[letterpaper, 12 pt, conference]{article}  
\usepackage{hyperref}

\usepackage{graphics} 
\usepackage{graphicx}
\usepackage{fullpage}
\usepackage{float}
\usepackage{amsfonts}
\usepackage{xspace}
\usepackage{enumitem}
\usepackage{booktabs}
\usepackage{makecell}
\setlist[description]{leftmargin=\parindent,labelindent=\parindent}

\newcommand{\makecellspaced}[1]{\vspace{0.2cm} \makecell[l]{#1}}

\newcommand{\RA}{\emph{RA}\xspace}
\newcommand{\ra}{\RA}
\newcommand{\PRV}{\emph{Prover}\xspace}
\newcommand{\prv}{\PRV}
\newcommand{\VRF}{\emph{Verifier}\xspace}
\newcommand{\vrf}{\VRF}
\newcommand{\CFA}{\emph{CFA}\xspace}
\title{\LARGE \bf
Remote Attestation: A Literature Review
}

\author{\makecell{Alexander Sprog{\o} Banks \\ alsb@itu.dk} \hspace{1pt} \makecell{Marek Kisiel\\ maki@itu.dk} \hspace{1pt} \makecell{Philip Korsholm \\ phko@itu.dk} \\
MSc Computer Science \\
IT University of Copenhagen \\
}
\date{December 2020}

\begin{document}

\maketitle

\begin{abstract}
With the rising number of IoT devices, the security of such devices becomes increasingly important. Remote attestation (\RA) is a distinct security service that allows a remote \vrf to reason about the state of an untrusted remote \prv (device). 

Paradigms of remote attestation span from exclusively software, in software-based attestation, to exclusively hardware-based. In between the extremes are hybrid attestation that utilize the enhanced security of secure hardware components in combination with the lower cost of purely software based implementations.

Traditional remote attestation protocols are concerned with reasoning about the state of a \prv. However, extensions to remote attestation also exist, such as code updates, device resets, erasure and attestation of the device's run-time state. 
Furthermore, as interconnected IoT devices are becoming increasingly more popular, so is the need for attestation of device swarms. 

We will describe and evaluate the state-of-the-art for remote attestation, which covers singular attestation of devices as well as newer research in the area of formally verified \RA protocols, swarm attestation and control-flow attestation. 

\end{abstract}

\newpage
\setcounter{tocdepth}{2} 
\tableofcontents
\newpage

\section{Introduction}
In recent years, the number of embedded systems, cyber-physical systems and Internet-of-Things (IoT) devices has increased significantly. Their growing presence intrudes many aspects of daily life, such as households, offices, and factories. Connecting these devices to the internet provides many benefits; however, it also expands the attack surface for an adversary. 

As users are becoming more dependent on the "smart" devices, their security becomes increasingly important.
In the context of actuation-capable devices, malware can impact security and safety, e.g., as demonstrated by Stuxnet\cite{stuxnet}. Furthermore, for sensing devices, malware may turn vulnerable devices into zombies that can become sources for distributed denial-of-service (DDoS) attacks\cite{vrased}, as demonstrated with the Mirai botnet\cite{mirai}. 

Unfortunately, security is typically not a key priority for low-end device manufacturers due to cost, size and power constraints. Therefore it is unrealistic for these devices to prevent attacks. Since it is difficult to prevent an adversary from compromising the device, we can instead try to check if the device has been infected or not. 
This requires Remote Attestation (\RA): "\textit{A distinct security service allowing a trusted party, called verifier, to validate or reason about the internal state (including memory and storage) of a remote untrusted, possibly infected with malware, party called the prover}"\cite{vrased}. 

Remote attestation can be extended to allow for remote code updates. This can be used to securely update software running on the device, reset the infected device, or erase the device. Furthermore, remote attestation typically considers a single \vrf and a single \prv. However, protocols that enable attestation of multiple \prv devices (swarm attestation) has in recent years become an active area of research. 

The goal of this report is to provide an overview of the state-of-the-art by critically evaluating remote attestation protocols. 
Furthermore, this report should provide an overview of what remote attestation is, how it can be accomplished and what an adversary may look like.

Throughout the report we will use a running example to describe the aspects of remote attestation. We will describe each facet from the perspective of a small internet connected IoT device. 
When an example requires changes to the device, such as hardware or software modifications, these requirements will also be described in context.
Furthermore, we may provide additional context to elaborate on certain aspects of the IoT device, e.g. provide a concrete use case to describe importance of certain requirements.

The report will describe the different paradigms of remote attestation (section \ref{sec:types_of_ra}), the typical adversary (section \ref{sec:adv_model}), the properties of remote attestation (section \ref{sec:properties}), what security architectures are (section \ref{sec:sec_architectures}), what remote attestation is (section \ref{sec:remote_attestation}), evaluate various remote attestation protocols (section \ref{sec:evaluation}) and lastly conclude on the state-of-the-art for remote attestation (section \ref{sec:conclusion}).

\subsection*{Use cases}
Remote attestation can be used on devices ranging from small legacy devices\cite{pioneer}, to large computationally capable devices\cite{opera}. Some remote attestation frameworks target larger devices that have the computational power to use asymmetric cryptographic primitives, whereas others rely on symmetric primitives that are not as computationally heavy.

Remote attestation can be used for attesting devices in many different scenarios. In hospital equipment remote attestation can be used to ensure that critical devices connected to the internet are not compromised and carrying malware\cite{sancus}.

Furthermore, remote attestation can be used in smart home scenarios to ensure the integrity of sensors and smart devices, thus ensuring that there are no false measurements created by an attacker and no secret data is leaked from the smart devices\cite{knox}.


\subsection*{Assumptions}
\label{sec:assumptions}
Some common assumptions made by remote attestation protocols are:

\begin{description}[style=nextline]
    \item[Physical adversary attacks are being disregarded] In \RA a general assumption is that no physical adversaries are present. This is because it is infeasible to defend against physical attacks in most scenarios.
    
    \item[Deployment of devices]
    In \ra it is assumed that the IoT device comes with pre-installed security architecture and attestation code.
    
    \item[Secret key distribution]
    It is assumed that the IoT device comes with a pre-installed shared key, \emph{k}, known by both the \vrf and \prv.
    
    \item[Cryptographic primitives]
    It is assumed to exist on the device and enable secure communication between the \prv and \vrf, or any other \vrf in a swarm. However, the specific implementation is not specified as many IoT devices have limited computational power, which limits the use of public-key-infrastructure.  
\end{description}


\newpage
\section{Paradigms}
\label{sec:types_of_ra}
Remote attestation can be categorized into three overall approaches based on implementation: hardware-based, software-based and hybrid remote attestation. Although not a paradigm, security architectures provide the foundation of many \RA protocols, and will be described initially in this section. Lastly the paradigm of swarm attestation will be introduced.

\subsection*{Security Architectures}
A security architecture provides trusted computing functionality for low-end IoT devices. The architecture should provide a number of security guarantees that provides a foundation for remote attestation, e.g. key-protection and atomic execution. These provided properties will be formalized in section \ref{sec:props_sec_arch}. 

\subsection*{Hardware-based attestation}
\noindent
Hardware-based remote attestation leverage physical chips and modules to achieve remote attestation. These modules include Trusted Platform Modules (TPMs) which are hardware modules that enables secure storage and computation, and dedicated processor architectures such as Intel SGX and ARM TrustZone.

\subsection*{Software-based attestation}

Software-based remote attestation does not rely on any hardware to perform remote attestation. The attestation program is run from memory to validate the software state of the system. These attestation protocols often rely on strong assumptions about the environment, such as the \prv not communicating with other devices during the attestation protocol and that communication is limited to one-hop.

\subsection*{Hybrid attestation}

Hybrid remote attestation is a hardware/software co-design that is based on a minimal trust anchor. The purpose is to combine the security guarantees of the hardware attestation approaches with the lower cost of software attestation to address the issues with software based remote attestation. These protocols use a minimal hardware footprint combined with software on the devices to provide a the guarantees required for remote attestation.

\subsection*{Swarm attestation}

Swarm attestation is a paradigm where there are many \emph{provers} instead of one. These \emph{provers} are connected by a network to allow communication between devices. Often, the devices are connected in a specific network topology to improve the performance or the fault tolerance of the network. The goal of swarm attestation is to attest the network of devices more efficiently than attesting each device individually.

\section{Adversary Model}
\label{sec:adv_model}
The general goal of an adversary may vary depending on the type of adversary and their motivation. However in an \RA setting the goal of the adversary is to remain undetected during an attestation while maintaining presence on the device. The means to do so vary with the type of adversary as they have different capabilities.
Five different types of adversaries with various levels of power and differing goals will be described\cite{invited}: 

\begin{description}[style=nextline]
    \item[Remote Adversary]
    The goal of a remote adversary is to remotely infect the \prv with malware\cite{invited}.
    A remote adversary is not on the network of devices, however it still has capabilities to infect the device from afar. This type of adversary is the most common, and can impact large networks of devices, e.g. the Mirai botnet\cite{mirai}.
    
    \item[Local Adversary]
    A local adversary is sufficiently near the \prv to be capable of eavesdropping on, and interfering with, the \emph{prover’s} communication. A local adversary is most likely on the network of devices and has strong computational capabilities\cite{invited}.
    
    \item[Physical Non-intrusive Adversary]
    A physical non-intrusive adversary is physically even closer to the \prv, so as to be capable of mounting side-channel attacks\cite{invited}.
    
    \item[Stealthy Physical Intrusive Adversary]
    A stealthy physical intrusive adversary can capture the \prv and attempt to physically extract any information (including secrets) stored thereupon\cite{invited}.
    
    \item[Physical Intrusive Adversary] 
    A physical intrusive adversary can physically capture the \prv. Furthermore, it can modify its state and/or hardware components (e.g., introduce additional memory)\cite{invited}.
    
\end{description}

As mentioned in section \ref{sec:assumptions} remote attestation protocols typically disregard physical adversaries, i.e. non-intrusive, stealthy and intrusive, as it is infeasible to defend against at protocol level.
However, it is possible to defend against physical adversaries by covering the CPU in tamper-resistant coating, as well as using anomaly detection and internal power regulators to prevent side-channel key leakage\cite{hydra}.

The most common attacks employed by remote and local adversaries are TOCTTOU and denial-of-service attacks, which will described below.

\subsection*{TOCTTOU Attacks}
Time-Of-Check-To-Time-Of-Use (TOCTTOU) attacks are some of the most common attacks against remote attestation protocols\cite{trap}. An adversary may record the correct code image after compromising a node. When the compromised node receives an attestation challenge request, the adversary returns the previously recorded correct code image (time-of-check). However, the executed code is different from the code checked (time-of-use) \cite{trap}. 

A TOCTTOU attack is especially prevalent when loading code from low-speed storage to high-speed memory, i.e. FLASH to RAM or cache\cite{minimalist}. An adversary could modify the attestation code after it is loaded into RAM, but before it is executed\cite{minimalist}.

The countermeasure commonly used against TOCTTOU is the use of nonces\cite{minimalist}. The \vrf will generate a nonce at attestation time which is sent to the \prv. When the \prv computes the attestation result, the nonce will be included in the digest and thus affect the result of the attestation. When the \vrf checks the attestation result the nonce will be factored in to ensure that the attestation is computed in response to the attestation request.


In a setting where IoT device is measuring data which reports back information to a \vrf. If the \vrf doesn't send a challenge or nonce in the attestation request, a remote or local adversary can forge and send an attestation request to the device before infecting it. When the device is then infected, if an attestation request is received the result of the previous attestation request can be forwarded to the \vrf. Because there is no check using nonce's the attestation response will seem valid to the \vrf.

\subsection*{DOS attacks}
Any device with network traffic is faced with the challenge of mitigating denial-of-service (DOS) attacks.
Remote attestation requests provide adversaries with another measure of launching a DoS attack. If an attacker can  send an attestation request to a device, that device will waste a significant amount of computational power computing a proof that the software on the device is benign. Given that IoT devices don't necessarily have large processors, this can affect the service provided by the device. A countermeasure for DoS attacks is to authenticate messages, this ensures the \prv does not process invalid attestation requests\cite{seed}. However, a local adversary may eavesdrop on the communication and use a replay attacks to circumvent this countermeasure\cite{seed}. The common countermeasure against replay attacks is the inclusion of nonces which is also prevalent in the domain of remote attestation.

The goal of remote attestation is not to mitigate DoS attacks, as most protocols simply assume that the device is compromised if the device does not respond.
This is of course not always sufficient. In the context of critical infrastructure where uptime of devices is pivotal, such as a hospital or nuclear reactor. In these settings assuming a device is compromised and thus disregarding the information provided by it is not enough and other measures have to be taken.


\section{Properties of remote attestation \& security architectures}
\label{sec:properties}
This section will introduce the properties that will be used to describe and evaluate \RA protocols (singular and swarm) and security architectures. 
The properties to describe \RA protocols are derived from the papers Principles of Remote Attestation\cite{principles_of_remote_attestation} and A Minimalist  Approach  to  Remote Attestation\cite{minimalist}. Furthermore, we have created a set of properties to describe security architectures and swarm attestation based on the common properties described in the literature reviewed for security architectures and swarm attestation respectively.

\subsubsection*{Properties of remote attestation protocols}
\label{sec:properties_of_ra}
\begin{description}[style=nextline]
    \item[Fresh information] An attestation should reflect the state of the \prv at the time of attestation.
    \item[Comprehensive information] Attestation mechanisms should be capable of delivering information that allows the \vrf to reason about the state of the \prv.
    \item[Trustworthy mechanism] \vrf should be able to reason that information received from \prv, is correct even in the presence of an active adversary.
    \item[Exclusive access] Only the \emph{prover's} attestation mechanism should have read access to the secret \textit{k}. No other process or device should be allowed to read the value of secret \textit{k}.
    \item[No leaks] The attestation mechanism should not leak any information that allows an adversary to reason about the secret \textit{k}.
    \item[Immutability] The attestation mechanism cannot be modified by an adversary with local or remote access to the device.  
    
    \item[Atomic execution] Execution of the attestation mechanism cannot be interrupted by any action invoked on the device. 
    \item[Controlled invocation] The attestation mechanism must only be invoked from its intended entry point.
\end{description}

\subsubsection*{Properties of swarm attestation protocols}
\label{sec:properties_of_swarm_ra}
A different set of properties will be used to describe swarm attestation protocols. These protocols are still reliant on the properties mentioned above, however the primary concern is not on those properties. Instead the protocols will be described based on the interaction of devices and what properties this interaction provide. The properties are as follows:

\begin{description}[style=nextline]
    \item[Efficiency] The attestation protocol must be more efficient than attesting each device one at a time\cite{seda}.
    \item[Aggregation of results] Which mechanism is used to gather the results of the individual attestations.
    \item[Attestation topology] The structure which is used to traverse the entire swarm e.g. using a spanning tree or a publish subscribe model.
    \item[Fault tolerance] Whether protocol is reliant on the swarm being stable or if devices can drop out of the network and rejoin later.
    \item[Control flow attestation] Indicates if the protocol supports control flow attestation (section \ref{sec:cfa}).
\end{description}




\subsubsection*{Properties of a Security Architecture}
\label{sec:props_sec_arch}
Additionally, a set of properties describing security architectures will be presented. These properties are compiled from several papers on different security architectures and will be used to describe and evaluate each of the architectures\cite{trustlite, smart, tytan, vrased, hydra, sancus, sancus2}.
\begin{description}[style=nextline]
    \item[Key Protection] A mechanism that allows some secret, \textit{k}, to remain secret from a local adversary as described in section~\ref{sec:adv_model}.
    \item[Immutability] Ensures that the data such as the attestation code or a secret key \emph{k} remains unchanged even in the presence of an adversary.
    \item[Atomic Execution] Ensures that code in the immutable code area cannot be interrupted during execution.
    \item[External Software] Allowing third-party vendors to develop and deploy in a secure environment.
    \item[Built-in \RA] The security architecture natively supports remote attestation.
\end{description}

\section{Security architectures}
\label{sec:sec_architectures}
Security architectures provide properties that many hybrid attestation protocols require. Often, a large subset of an \RA protocol's properties, as described in section \ref{sec:properties_of_ra}, are obtained via the underlying security architecture. Furthermore, as seen in section \ref{sec:props_sec_arch}, security architectures can be reasoned about from a set of properties as well.



Specialized hardware such as read-only-memory(ROM) is often used by security architectures to provide both immutability and key protection. Immutability is obtained by using ROM which is write protected and thus ensures the code cannot be changed\cite{trustlite, sancus, smart, hydra}. 
Key protection can be obtained using ROM, but with additional measures to ensure that only specific code has access to the key. These measures can be enforcing that the CPU instruction pointer points to a trusted code region in ROM\cite{smart} or using a memory protection unit (MPU)\cite{hydra, trustlite}. All security architectures that support Built-in \RA typically provide both key protection and immutability.

Some security architectures allow for trusted and isolated software execution environments. These environments provide third party software vendors with an environment in which they can securely deploy code to extend and customize the functionality of the device\cite{trustlite, sancus}. The isolated software components may need to communicate with one another, which must be done in a secure way, also known as Inter-Process Communication (IPC). IPC can be realized with message queues in the operating system (OS). The OS will notify when new tasks are available in the message queue\cite{trustlite}. To ensure the security of the communication, an MPU can be used\cite{trustlite} to ensure that protected code is not invoked from an invalid point(given that a single entrypoint to a deployed software module is provided). Alternatively, communication among these environments can be obtained using MACs and key-derivation functions. In this case the security architecture can allow module to communicate if they declare, via MACs, the key of the module with which they wish to communicate.\cite{sancus}.

Atomic execution does not allow any interrupts of specific code in memory, such as attestation code. The goal of this is, among other things, to ensure that no intermediate values in the CPU registers are leaked and to prevent return-oriented programming (ROP) attacks\cite{minimalist}.
Atomic execution can be obtained either by modifying the kernel to disable interrupts at certain times or by the attestation code itself disabling interrupts.
The security kernel seL4\cite{hydra} provide the functionality of ensuring that code will not be interrupted, as it is possible to boot up an applications with different priority levels. This ensures that no program executed with lower priority can not access, modify or interrupt the protected code.

Alternatively, atomic execution can be implemented using an execution-aware MPU (EA-MPU) and program counter rules. In this way if the attestation protocol is accessed using an invalid entry sequence, or an interrupt is raised, the MPU will reset the device\cite{vrased}. This reset allows an adversary to conduct a DoS attack on the attestation protocol by continually resetting the device. However, in many protocols devices who don't respond are considered compromised. \ra protocols which have an update implementation will have the possibility of updating the state of compromised device to its original state.

An EA-MPU can also be used to allow interrupts of secure code by controlling what the interrupts affect\cite{trustlite, tytan}.
Depending on the use-case interrupts may be necessary, e.g. by allowing a network package to be received or if the device must perform a system-critical task\cite{erasmus}.









\section{Remote attestation}
\label{sec:remote_attestation}





\begin{figure}[H]
    \centering
    \includegraphics{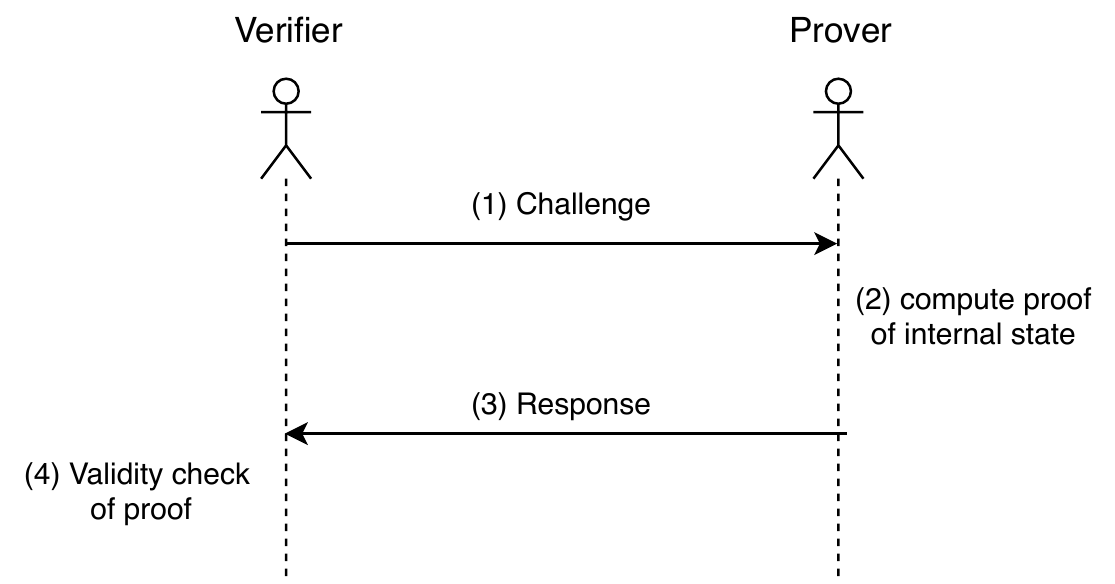}
    \caption{Challenge-response protocol}
    \label{fig:challenge-reponse}
\end{figure}

This section will describe \RA and how it can be achieved in various settings. It will cover singular attestation, which is attestation of a single \vrf and a single \prv, and swarm attestation.
Furthermore, extensions to \RA will be described, i.e. code updates, erasure, and reset, as well as shuffled measurements and control flow attestation. 
A common way to implement \RA is a challenge-response protocol. Such a protocol can be realised in four steps as can be seen in Figure~\ref{fig:challenge-reponse}. 

\begin{enumerate}
    \item Initially the \vrf generates a challenge e.g. a random bitstring.
    \item The \vrf sends the challenge to the \prv.
    \item The \prv calculates a proof of its local state and forwards it to the \vrf.
    \item Lastly the \vrf validates the response to the challenge.
\end{enumerate}

During step 2 of the attestation the \prv will compute a digest of the memory region being attested. This is common in \RA, and requires the \vrf to have knowledge of the possible memory states that the \prv can enter. This interaction can be formalized as a 3-tuple of operations that the \prv and \vrf execute together: (Request, Attest, Verify).

\begin{description}[style=nextline]
    \item[Request()] The \vrf generates a random challenge $Chal$ is generated and sent to the \prv. 
    \item[Attest$(Chal)$] The \prv self attests and includes the challenge in the digest \emph{H}. The digest is then sent to the \vrf.
    \item[Verify$(H, Chal)$] The \vrf checks the validity of the response generated by the \prv. The knowledge of the valid internal states of the \prv is used in this step to verify the digest.
\end{description}

The implementation of these steps vary between protocols. 
Some protocols also deviate from the challenge-response structure, and instead opt for a non-interactive protocol. In this way the \vrf doesn't send a challenge, instead the \prv will at certain intervals send an attestation to the \vrf.

\subsection{Singular attestation}
In section~\ref{sec:types_of_ra} we introduced various paradigms in \RA, i.e. software-based, hardware-based and hybrid remote attestation. In this section we will elaborate on how each of these can be achieved, their limitations, assumptions, and provide a holistic overview of how each paradigm work.

\subsubsection{Software-based remote attestation}
Software-based attestation is a type of remote attestation protocol which does not rely on access to specialized hardware. Since software-based \ra does not require any hardware modifications, the implementation and deployment is cheap and uncomplicated due to the lack of specialized hardware. However, not relying on special hardware limits the security guarantees which can be obtained. 
Software-based \ra relies on computing a checksum to achieve a \emph{Trustworthy mechanism}. During the execution of \emph{Attest} the device will first calculate a checksum of the attestation code and then attest the software on the device. The checksum will also be sent to the \vrf who can validate the integrity of the attestation code and the attestation.

Software-based \RA is often based on the computation time of a checksum\cite{scuba, swatt, pioneer}. Since the checksum time is dependent on the configuration of the individual device, an adversary with more computational power might have a possibility to exploit the timing property.
Software-based attestation protocols guard against these attacks by relying on the assumptions that adversaries cannot collude with other devices, that the checksum and attestation code cannot be optimized further and that the attestation cannot be parallelized. 
The assumptions made by the attestation protocols defends against the adversaries. However if collusion is possible a local adversary would have the advantage of lower latency's in the network. If the local adversary has stronger computational capability the attestation could be computed on a faster machine thus breaking the attestation. Software-based \RA rely on strong assumptions of adversarial capabilities and only work if the \vrf communicates directly to the \prv. This direct communication is also referred to as one-hop communication\cite{vrased}.

Despite these restrictions, software-based solutions have been adopted on several embedded systems such as SBAP~\cite{sbap}.
SBAP uses remote attestation to verify the integrity of peripheral firmware. Which is implemented and evaluated on an Apple Aluminium Keyboard~\cite{sbap}. Software based \RA can be used in this case as attestation of peripheral is within one-hop on a local network.

In our running example implementing software based remote attestation on the device would require no modifications to the device. The attestation code could be loaded onto the device before deployment or it could be retrofitted after the device is deployed. However to allow software based attestation to work the IoT device(\prv) cannot be deployed longer than one hop away from the \vrf. If the device was not at one-hop distance the timing checks between \prv and \vrf would not be reliable. 


\subsubsection{Hardware-based remote attestation}
With hardware-based \ra it is possible to achieve the highest security guarantees, providing the possibility of being implemented in critical scenarios. Hardware-based \RA may rely on the \textit{Trusted Platform Modules} (TPMs), which is a co-processor designed to protect cryptographic keys and record the software state of a computing platform. The objective of the TPM, is to provide a hardware-based root of trust for the computing system~\cite{invited}. This is achieved by using special-purpose Platform Configuration Registers (PCRs). Each of these registers can store a cryptographic hash. Overwriting a register in a TPM is not allowed, but extending the register value is allowed. A register can be extended by hashing the currently stored value with another hash value~\cite{invited}:

$$
 PCR[0] \leftarrow H(PCR[0]\; |\; h)
$$

Here H is some hash function, h is some cryptographic hash and $|$ is concatenation. 
When a device performs a measured boot, each binary in the boot chain will perform a measurement (compute the hash) of the the subsequent binary, record it in the measurement log, and finally extend it into the PCRs~\cite{invited}. To validate the integrity of the measurement log, we can compare the sequence of measured values to the current PCR values. The PCRs provide an easily verifiable chain of values, which were accumulated since the platform was last reset.

Since the PCR values are a representation of the system’s state, they can be used as attestation evidence~\cite{invited}. Thus, if there is any discrepancy between the measurement log and the PCRs, the integrity has been compromised. 

Attacks on hardware based attestation are considered infeasible since the base components used for hardware based attestation is considered secure hardware. The hardware is dedicated to security purposes and cannot be attacked by regular physical attacks such as side-channel attacks. This is the most secure type of attestation but also the most expensive since dedicated secure hardware is expensive to deploy on many devices\cite{invited}. 

Hardware-based \RA has strong security guarantees and is useful in scenarios where the expense of the hardware is less important than the function of the device. In a hospital setting hardware-based remote attestation can be relevant due to the importance of ensuring that medical devices function correctly.

\subsubsection{Hybrid remote attestation}
As previously mentioned in section \ref{sec:types_of_ra} hybrid attestation utilizes hardware/software co-design to create attestation protocol that provide stronger guarantees than software based attestation but is cheaper than hardware based attestation. 

For hybrid attestation the attestation code needs to be located in an immutable storage, that will allow the attestation code to remain unchanged. While also remaining unchanged, the access to the secret key \emph{k} for attestation should be limited to only the attestation code.
Hybrid attestation schemes will often rely on a security architecture to achieve these guarantees and to obtain the properties described in section~\ref{sec:properties_of_ra}. In these cases the security architecture relies on the hardware modifications on the device for these guarantees along with the implementation of the architecture itself.
The underlying security architecture might provide ROM and MPU, where the ROM would act as storage for secret key \emph{k} and the attestation code\cite{vrased}. Storing the information in ROM provides immutability property of \ra of immutability (see section \ref{sec:properties_of_ra}). Whereas the MPU, by the use of access control, enable only the authorized process in accessing the memory of ROM, so that no information about the secret key would be leaked, providing the exclusive access property of \ra(see section \ref{sec:properties_of_ra}). 

The hybrid attestation approaches relies on the properties provided by the security architectures. Local and remote adversaries who only operate on the software are unable to break attestation due to the key protection and immutability of the attestation code. Furthermore the attestation protocols rely on the security of HMAC algorithms and similar cryptographic functions. If either of these are violated the attestation will be vulnerable. Physical adversaries can use side-channel attacks to extract secrets from read-only memory, however as mentioned in section~\ref{sec:adv_model} this issue is not something that is considered in remote attestation.



Many of the hybrid attestation protocols use a challenge-response approach to attestation. There are some downsides to this approach, mainly that the device either has to complete the task it's currently performing before attesting, or it has to stop executing while it executes the attestation. Having to stop executing can in some critical cases be relevant. In hospital equipment devices may not be able to stop execution during operation, and thus a challenge based protocol is not feasible. An approach to combat this problem is to continuously attest memory when the device is not in critical operation. In this type of protocol the \emph{Request} will be self initiated and use a shared secret key \emph{k} stored on the device for attestation as well as a timestamp. These attestations can then be stored in a circular memory buffer to limit memory consumption. When the \vrf wishes to verify the internal state of a device, it sends an attestation request and \prv responds with the content of this buffer. In this way attestation doesn't happen on demand, instead when the \prv has the available resources to attest. This method of attesting provides the benefit that the \vrf can know if the \prv was compromised between attestation requests\cite{erasmus}.






\subsection{Swarm attestation}

The goal of swarm attestation, also known as collective attestation, is to attest a network of devices in a way that is faster than attesting each device individually\cite{seda}. This property will be referred to as efficiency, as mentioned in section~\ref{sec:properties_of_ra}.
This section will introduce the concept  of scalable \RA protocols, that work for a swarm of devices, and describe how swarm attestation can be accomplished.

Swarm attestation is a type of \RA that, instead of describing how attestation of a single device is conducted, describes how attestation of many devices is orchestrated. These protocols are usually not dependent on the underlying \RA paradigm i.e. software, hardware, hybrid. Most swarm protocols are implemented in a hybrid setting, but there are examples of hardware-based swarm attestation schemes as well e.g. OPERA\cite{opera} and TRAP\cite{trap}. 

In swarm settings the \emph{Request} part of the attestation is propagated, either with a fresh challenge or reusing the same. The node at which the \emph{Verify} step is conducted varies between protocols. Some protocols use distributed verification, which means that each \prv also act as a \vrf. In the case of distributed verification \emph{Verify} is executed at each node and the results are propagated. If verification is not distributed the results are aggregated and a \vrf can eventually validate the result of the attestation.

Swarm attestation protocols use various topologies for propagating attestation requests, to accommodate how disruptive the network is, i.e. devices joining and leaving the network. 
The devices in a swarm can be connected in a tree structure (e.g. spanning and binary trees), which allows for recursively aggregating attestation requests throughout the network\cite{seda,sana,lisa}. The issue with this structure, is that it becomes expensive to rebuild the spanning tree in a dynamic network setting, i.e. a tree network topology provides poor fault tolerance\cite{pasta}. 
Due to the low fault tolerance of a tree structure, a mesh network topology can be used to accommodate this. It allows for a dynamic network where connections among devices frequently change\cite{pasta}. 

Devices in the network can communicate with each other in various secure ways. A popular approach is to use MAC and signature schemes\cite{seda, pasta, salad} to ensure integrity of the attestation results. However, there are also other alternatives using asymmetric encryption\cite{radis}.

\subsection{Extensions to remote attestation} \label{sec:extensions}
Remote attestation on its own only provides functionality to validate the internal state of devices. If a device is compromised there are two options for the \vrf; blacklisting or updating/resetting it. For some use cases blacklisting a device is acceptable, otherwise updating or resetting a device is desirable. Extensions to remote attestation schemes such as update, reset and erasure will be described in this section, as well as how each of them are achieved. Furthermore we will describe a technique for measuring the software state called shuffled measurements which attests random portions of memory. Lastly, we will describe control-flow attestation, which is attesting the run-time state of a binary. 

\subsubsection*{Update} 
Updating a device means updating a part of the software present on the device. There are two reasons for wanting to update the software of a remote device 1) the device has been compromised and the software should be patched and the intruder shut off from the system 2) a software update is released by the vendor, either because a vulnerability has been found but not yet exploited and thus needs to be patched, or a software update is released with new or changed functionality.
In the process of updating a device \RA can ensure that the update sent to the device is correctly installed. To verify an update, a scheme consisting of a three-tuple can describe the operations: (Request, Install, Verify)\cite{pure}.

\begin{description}[style=nextline]
    \item[Request$(Chal, S)$] Is initiated from the \vrf. A challenge $Chal$ is generated and sent to the \prv along with $S$ that is the update for the \prv to install.
    \item[Install$(Chal,S)$] The \prv installs \emph{S} and computes the response to the challenge \emph{Chal}, which is sent to the \vrf as \emph{H}.
    \item[Verify$(H, Chal, S)$] Lastly the \vrf can verify the response to the challenge that indicates whether or not the installation was successful.
\end{description}

\noindent
This scheme extends the functionality of \RA protocols described previously, and uses \RA as a subroutine after the installation has occurred. During \emph{Install} the \prv computes the response to the challenge, which in this case is an attestation of the update. This allows the \vrf to simply use the existing \emph{Verify} routine already implemented in the \RA protocol.

To install a software update two methodologies have been implemented in the \RA protocols investigated: Using a secure boot-loader\cite{trap} to load memory, and sending memory snippets using the same communication channels that the \RA protocol uses\cite{pure, scuba}. 

When using a boot-loader the updates for programs are again sent from the \vrf during reboot of a device. When the device boots up it will listen for incoming updates. While an update is being received, it is written into external flash memory, and when the entire update has been written into memory it is copied to data memory where it can run\cite{trap}. 

When sending memory snippets using existing communication channels, \VRF inspects the results of the initial attestation to determine which memory regions needs patching. If the initial attestation doesn't reveal any invalid state of \PRV and update might not be needed. If an update is required \VRF constructs and sends these patches to the device\cite{scuba}. On the device the memory snippets are placed at the correct spots in memory and the \PRV does an additional attestation of the memory to validate to the \VRF that it has been updated\cite{pure}. This approach can also be used if the security architecture of the device supports running isolated software modules. A module can be created, and the existing modules extended, to allow modules to update each other\cite{trustlite, sancus}.

In our running example of an IoT device update functionality can prove very useful. If a new zero-day vulnerability is disclosed and patched in the software that the device is running, updating the device is desirable. However, if the device does not support code updates it will remain vulnerable, which in turn will allow adversaries to gain access to the device. In this case there is no measures that can be taken unless you physically remove the device from the network to patch it or disregard the information obtained by the device. 


%
%
%
\subsubsection*{Reset}
\RA might benefit from being able to reset execution on a device remotely for a number of reasons. In a general purpose scenario a reset can be used to restart software execution flow. If \PRV has reached an unexpected state a reset can restore it to a valid state. After an update a reset can be used to ensure that new software is executed from a valid state. If software is updated by inserting code into memory on run-time a reset might be required to avoid invalid control flows\cite{pure}.

To reset and obtain a proof of reset, there are three steps to follow by \PRV 
To reset and obtain a proof of reset, \PRV must do the following after receiving a reset request: 1) a reset request contains a challenge. Therefore, it must attest the reset method using the challenge value. The result of this attestation is stored in a memory address that will not be reset. 2) The memory and program counter is reset 3) The result of attesting the reset function is returned to the \VRF as the proof of reset\cite{pure}.

\subsubsection*{Erasure}
Erasure means to erase program information from memory. To be able to erase programs from memory is useful if a device is being decommissioned or it changes ownership. Furthermore it can be used in conjunction with an update(mentioned previously), to clean the memory before installing updated programs\cite{pure}.

Memory Erasure can be viewed as a special case of updating a device where instead of new bytes containing updates to a program, the bytes sent from \vrf are 0 bytes. In this way the area that is being "updated" is being erased. 

Viewing erasure as a case of update means that memory erasure is a property that follows from update, and a proof of erasure would be identical to update.

%
%
\subsubsection*{Shuffled measurements}
Shuffled measurements is an extension to regular attestation techniques used to defend against roving, or relocating, malware \cite{smarm}. 
Shuffled measurements means attesting memory in a random order to detect malware that use relocation to avoid detection.
Using shuffled measurements requires awareness about the knowledge of potential adversaries. The literature distinguishes between three different adversaries based on their knowledge. These adversaries are categorized as non-physical adversaries, though not specifically local or remote. 
The three levels of knowledge proposed are Knowledge of Future Volume(KFV), Knowledge of Future Coverage(KFC) and Knowledge of Future Order(KFO). KFV means knowing the size of the memory yet to be attested, this can be achieved by measuring the time since attestation was initiated and dividing by the average time to attest a block of memory. KFV is the least powerful knowledge that an adversary can obtain. KFC is an extension of KFV, with KFC the adversary is aware of how many blocks are left to attest, it also knows which blocks have been attested. An adversary with KFC knowledge is more powerful than one with KFV knowledge since if it evades detection in the first block of attestation, the malware can move to the block which has been attested. Knowledge of Future Volumes can be obtained if an adversary can distinguished attested blocks of memory from blocks which have not been attested. Lastly, an adversary with KFO knowledge is the most powerful, Knowledge of Future Order means the adversary knows which blocks will be attested next. This means an adversary always can avoid detection. Knowledge of Future Order can be obtained if the attestation code insecurely stores which blocks are attested next. 

\subsubsection*{Control Flow Attestation} \label{sec:cfa}
\label{sec:control_flow_attest}
The previously described remote attestation protocols can only ensure integrity of the binaries, and not the execution and run-time. This doesn't provide a complete picture of the internal state of a device, because the behavior of a device at run-time can be changed without modifying the binaries. The solution to this problem is control flow attestation (\CFA).

We define \CFA as an extension to \RA protocols, as it provides additional security guarantees in regards to the software's run-time integrity. However, the implementations also result in larger computational overhead and more complex protocols\cite{cflat}.

\begin{figure}[H]
    \centering
    \includegraphics{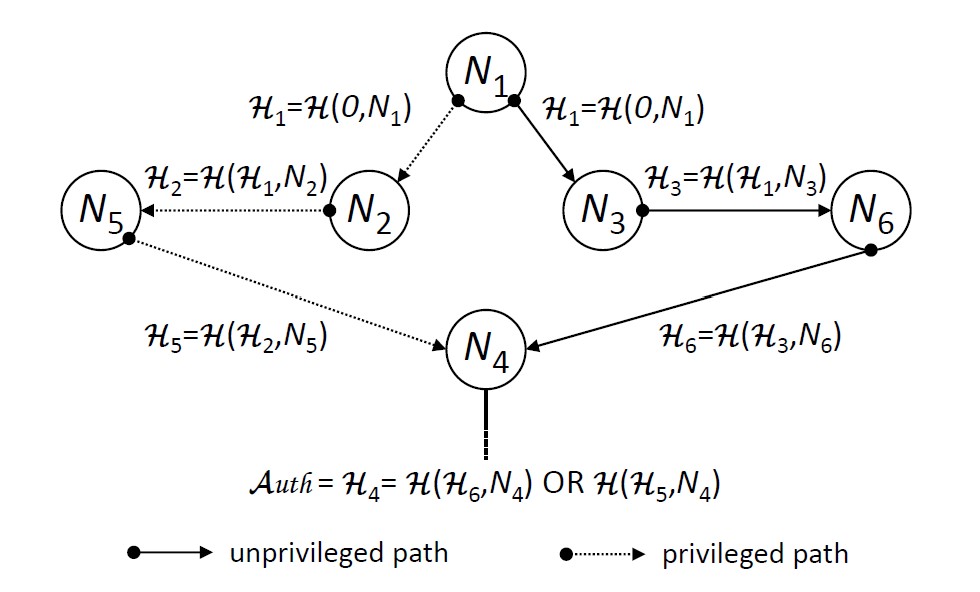}
    \caption{Control Flow Graph from C-FLAT\cite{cflat}}
    \label{fig:cfg}
\end{figure}

The goal of \CFA is to provide remote attestation of a device's run-time\cite{radis}. This can be done using a control flow graph (\emph{CFG}) where each node represents a block of assembly instructions. Each path in the \emph{CFG} represents a branching, which can be either privileged or unprivileged path as seen in Figure \ref{fig:cfg}. The need for control flow graphs decrease the complexity of binaries which can be attested. The size and complexity of the \emph{CFG} is linear proportional with the number of loops and recursive calls\cite{cflat}. An increase in the size of the \emph{CFG} will also increase in the amount of hashes that need to be stored on the \vrf. Therefore it is possible to define the loops and recursive as a standalone program in order to minimize the \emph{CFG}. 


Each device can calculate a hash based on each node's id, and the previous hash. This creates a hash chain with a limited set of valid hashes, as can be seen in figure~\ref{fig:cfg}, which the \VRF can validate at attestation time\cite{cflat}.

In regard to the running IoT example, consider that a local adversary subverts the control-flow of a device on a distributed network of devices. For context, assume a Smart Home setup with three devices: a camera, a controller and a door lock. Assume an adversary has compromised the controller.
To unlock the door the camera must take an image of the resident of a home (1), the controller checks that the image is valid (2), and finally the smart door lock unlocks (3).
Given the adversary has compromised the controller, it can modify the control-flow and successfully unlock the door.
If control-flow attestation was running individually on all devices, we would see that the controller was compromised, but that the door lock was in a legitimate state. 
Control-flow attestation protocols, running on every device, cannot detect the devices that have been influenced by the attacker and forced into an incorrect state\cite{radis}.



\newpage
\section{Evaluation} 
\label{sec:evaluation}


%


In this section we will evaluate singular and swarm \RA, as well as security architectures. The singular \RA frameworks will be evaluated in their respective approaches, i.e. software-based, hardware-based and hybrid.


\subsection{Security architectures}

In this section we will elaborate, evaluate and compare the three security architectures TrustLite, VRASED, and SANCUS.
We have chosen to highlight these, as we believe they are the most promising security architectures that we have investigated. 

Table \ref{tab:sec_arch_eval} shows the three security architectures horizontally, and the previously defined properties for security architectures vertically(section~\ref{sec:properties_of_ra}). 
The content of each cell describes how the given security architecture obtains the given property.


\begin{table}[H]
    \centering
    \begin{tabular}{@{}lcccc@{}}
    \toprule
         &  TrustLite & VRASED & SANCUS \\ \midrule
        Key Protection & EA-MPU & \makecell{ROM and PC \\ based memory control}\vspace{0.2cm} & \makecell{Protected storage area via \\ PC based memory control}\\ 
        Immutability &  PROM & ROM & \makecell{PC based \\ memory control} \\ 
        Atomic Execution & N/A & By reset and PC &  N/A \\ 
        External Software &  EA-MPU and IPC & N/A & Custom C-Compiler\\
        Built-in \RA & N/A & Formally Verified RA & N/A \\ \bottomrule 
    \end{tabular}
    \caption{Comparison of security architectures}
    \label{tab:sec_arch_eval}
\end{table}


\begin{description}
\item[SANCUS \cite{sancus}] was introduced in 2013 as a security architecture that enables third parties to develop and deploy software modules to IoT devices. SANCUS uses a customized C compiler to compile software modules from external vendors. This compiler ensures that all modules have a defined entrypoint and that all secret information is protected. 
To ensure the software modules are not tampered with, SANCUS uses \textit{program-counter (PC) based memory access control}. The access control provide isolation between software modules as well as isolation of keys distributed on the nodes. SANCUS uses secure linking for IPC by including the identity of the other software module in the current module. When executing the module, SANCUS allow the modules to communicate.



\item[TrustLite \cite{trustlite}] was introduced in 2014 by Intel as a security architecture that would allow hardware-enforced flexible and efficient software isolation (trustlet). This is achieved by using ROM and an execution-aware memory protection unit (EA-MPU). 
The EA-MPU allows inter-process communication (IPC) between different trustlets by defining the access rules for data and memory on the trustlet. Each trustlet has two entrypoints, \emph{call}, to invoke the trustlet from other trustlets(IPC), and \emph{continue}, to resume execution of the trustlet. Trustlet code cannot be invoked at different points than these.

\item[VRASED \cite{vrased}]
was introduced in 2019 both as a security architecture and remote attestation protocol. Compared to the previous security architectures, VRASED has an implementation of Remote Attestation as part of the security architecture. 
However, VRASED does not provide the possibility for external software vendors to deploy software on the device.
The attestation code and secret key will be stored in ROM in order to ensure that it is immutable.
VRASED, like SANCUS, uses PC based memory access control to ensure the key is protected, and that the attestation code cannot be modified or interrupted. It continually monitors the counter to ensure that the program doesn't engage in undefined behavior such as return-oriented programming.

\end{description}

SANCUS and TrustLite are more similar than VRASED since they provide many of the same functionalities. Whereas VRASED is a security architecture developed specifically for a formally verified remote attestation protocol.
Both SANCUS and TrustLite provide extendability with external software that can run in an isolated environment.

    

\subsection{Singular Attestation}
In this section we will evaluate several protocols from each of the previously described domains: hardware-based, software-based, and hybrid attestation. For each domain, a set of protocols will be described, and their properties, differences and use-cases evaluated.
These protocols present a subset of all remote attestation protocols, but are selected to cover different use cases and methods of achieving remote attestation.

\subsubsection{Software-based attestation}
Software-based \ra, can be used if remote attestation is retrofitted on devices after deployment, and adding physical components to devices is not an option. Software attestation relies on stronger assumptions than other types of remote attestation. 

All software-based remote attestation systems we have examined provide on demand attestation using a challenge-response protocol. They all rely on timing as well as the result of the attestation to gather information about the \prv. 

All software-based protocols achieve atomic execution by disabling interrupts on the operating system they are implemented on.
They do not have controlled invocation, meaning that an adversary could use return-oriented programming to attack the protocols, however they checksum the attestation code before the attestation is executed, thus any changes will be detected.

In table \ref{tab:software_eval} three different software-based remote attestation protocols are compared based on the properties from section~\ref{sec:properties_of_ra}. 
None of the protocols achieve the \textit{No leaks} and \textit{Exclusive access} properties. This is due to a lack of keys in software based remote attestation.

\begin{table}[H]
\centering
\begin{tabular}{@{}lllll@{}}
\toprule
      & SWATT & Pioneer & SCUBA \\ \midrule
Fresh information & On demand    & On demand & On demand \\
\makecellspaced{Comprehensive \\ information} & \makecell[l]{through timing \\ and attestation}    & \makecell[l]{through timing \\ and attestation}    & \makecell[l]{through timing \\ and attestation}    \\
\makecell[l]{Trustworthy \\ mechanism} & \makecell[l]{by assumption \\ memory- \\  content- \\ verification}    & \makecell[l]{Dynamic root \\ of trust using \\ checksum} & \makecell[l]{Untampered code \\ execution}    \\
No leaks & -    & -    & -     \\
Exclusive access &-    & - & - \\
Immutability & \makecell[l]{ - } 
& \makecell[l]{checksum of code \\ is computed \\ initially} &  \makecell[l]{checksum of code \\  is computed \\ initially} \\
Atomic execution & \makecell[l]{interrupts disabled} & \makecell[l]{interrupts disabled} &   \makecell[l]{interrupts disabled}  \\
\makecell[l]{Controlled \\ invocation} & - & - & - \\ \bottomrule
\end{tabular}

\caption{Software-based remote attestation protocols}
\label{tab:software_eval}
\end{table}


\begin{description}
    \item[SWATT \cite{swatt}] is a software attestation system for legacy devices, that also protects against roving malware. SWATT uses the challenge received in the challenge-response protocol as input to a pseudo-random generator (PRG). Based on the result of the PRG the \prv computes the checksum of random parts of memory. This approach is combined with timing checks, such that if an adversary redirects the checksum code to a different memory address or tries to change the content of what is on the address being examined the checksum function will slow down significantly. The Verifier will then know that an adversary has manipulated with the results. Due to the lack of immutability the adversary can change the memory-content verification function. However, SWATT assumes that a secure design will ensure the function fails if the memory content of the device doesn't match the expected content. SWATT can be used in existing networks to allow some degree of attestation in networks without dedicated hardware.
    
    \item[Pioneer \cite{pioneer}] also ensures the integrity of the attestation function by initially computing a checksum of the attestation function as well as the checksum code. This checksum also includes a challenge generated by the Verifier to ensure that the checksum cannot be precomputed. Pioneer relies on hash functions for attesting the memory, computing a hash of the relevant region to send the Verifier. 
    The checksum code is not parallelizable as well as optimal, meaning that no changes could be made to the code that would speed up the process and allow an adversary to manipulate what memory is attested.
    
    \item[SCUBA \cite{scuba}] relies on stronger assumptions than SWATT and Pioneer, assuming Read-Only memory to be present on the device. 
    SCUBA uses a public key infrastructure for communication between \prv and \vrf unlike the other protocols. 
    During the attestation a MAC is computed, however this doesn't use a shared secret, instead the attestation code uses the challenge received from the \vrf as an input for a pseudo-random generator, which outputs a key for the function. SCUBA achieves code updates on the device by registering if the attested code region has been modified. The \vrf sends code patches for certain memory regions to the \prv which the \prv will install.
\end{description}

\noindent

Although SCUBA categorizes itself as a software-based \RA protocol, we would argue that it fits more into hybrid \RA. This is because it relies on hardware assumptions, i.e. that the device has ROM\cite{scuba}. 
The verification function in SWATT, similar to Pioneer, is constructed so that any attempt to tamper with it will increase its running time. This is achieved without any hardware assumptions like SCUBA. 
However, both protocols require extensive knowledge of the underlying system, such as the CPU's model and clock speed, instruction set architecture etc.

Pioneer differentiates itself from SWATT, as it supports complex CPU architectures\cite{pioneer}. SWATT does not support CPU architectures with sophisticated features such as virtual memory and branch predictors\cite{swatt}. Furthermore, as SWATT checks the entire memory, its operating time on systems with large memories is also prohibitive\cite{pioneer}.

\subsubsection{Hardware-based attestation}

Hardware-based \RA relies on security guarantees entirely provided by hardware. In this section we evaluate three hardware-based \RA protocols. In table~\ref{tab:hardware_eval} we provide an overview of how each protocol obtains their respective security properties.
Atomic execution is marked as N/A, as interrupts are not applicable for any of the hardware-based \RA protocols. This is because no software is scheduled on the processor in a way that can leak information about secret keys or in other ways cause compromise to security.
Although table~\ref{tab:hardware_eval} may seem redundant, it highlights that hardware-based \RA essentially outsources all of its requirements to the underlying hardware, whether it be TPMs, Intel SGX or ARM TrustZone.

\begin{table}[H]
\centering

\begin{tabular}{@{}lllll@{}}
\toprule
      & TRAP & OPERA & KNOX \\ \midrule
Fresh information & On-demand    & On-Demand     & On-Demand    \\
Comprehensive information & TPM    & SGX     & TrustZone    \\
Trustworthy mechanism & TPM    & SGX     & TrustZone    \\
No leaks & TPM    & SGX     & TrustZone    \\
Exclusive access & TPM    & SGX     & TrustZone    \\
Immutability & TPM    & SGX     & TrustZone    \\
Atomic execution & N/A    & N/A     & N/A    \\
Controlled invocation & TPM    & SGX     & TrustZone    \\ \bottomrule
\end{tabular}

\caption{Hardware remote attestation protocols}
\label{tab:hardware_eval}
\end{table}

\begin{description}
    \item[TRAP \cite{trap}] is a hardware-based \RA protocol for wireless sensor networks (WSNs), using Trusted Platform Modules (TPMs). TRAP assumes each device(\prv) in the WSN is equipped with TPMs.
    The primary threat discussed in TRAP, is the Time-Of-Check-To-Time-Of-Use (TOCTTOU) attack, a common attack against \RA protocols. Communication between the \vrf and \prv is done using symmetric keys, which are shared before deployment of the devices. 
    
    \item[OPERA \cite{opera}] is a remote attestation protocol based on Intel SGX. OPERA seeks to improve openness, privacy and performance, which are some of the limitations of the Intel-centric attestation model. 
    The primary use case of OPERA is cloud computing. It can support centralized attestation as in the case of confidential cloud computing and privacy-preserving blockchains. Furthermore, OPERA can also be implemented to support decentralized attestation, in which enclave programs on the same SGX platform attest one another.
    As the root secrets of SGX platforms are controlled by Intel, OPERA must trust Intel to faithfully establish a chain of trust to identify SGX platforms and their protected software. The OPERA attestation model itself has been formally verified using ProVerif.
    
    \item[Knox \cite{knox}] is a platform available on Samsung smartphones, which allows for device attestation using ARM TrustZone. Knox was initially developed by Samsung to ensure that their smartphones' remain safe from a booting state to a running state. However, Knox also provides an API to check the current safety status including an attestation function. To perform remote attestation the Knox platform requires a third party server known as a Knox Attestation Server. Attestation using Knox is initiated by the \vrf by requesting a nonce from the attestation server that is then forwarded to the \prv in an attestation request. The \prv will then self attest and send the result to the \vrf. To verify the result of the attestation the \vrf forwards the result of the attestation to the attestation server. The attestation server validates the result and sends the result to the \vrf.

\end{description}

All of the presented hardware-based \RA protocols have somewhat specific use cases. Knox is \RA for Samsung smartphones, OPERA is \RA for Intel SGX enclaves and TRAP is \RA for hardware based Wireless Sensor Networks (WSNs). However, when using Knox and OPERA, you are forced to trust a third party. In Knox you must trust the security of the Samsung developed Knox Platform~\cite{knox}, and with OPERA you must trust the security of the Intel SGX platform~\cite{opera}. TRAP distinguishes itself from the other two protocol in this sense, as no third party is trusted. Furthermore, TRAP provides strong security guarantees as its root of trust in based on TPMs. However, equipping each device with TPMs in a WSN may be somewhat costly.

\subsubsection{Hybrid attestation}\label{eval_hybrid}
We have chosen to compare the \RA protocols VRASED~\cite{vrased}, SeED~\cite{seed}, SMARM~\cite{smarm} and Erasmus~\cite{erasmus}. These protocols represent distinct approaches to hybrid \RA, for various use cases. In table~\ref{tab:hybrid_eval}, the protocols are described in terms of their properties and how they are obtained. X denotes that the property is obtained by delegation to an underlying security architecture.


\begin{table}[H]
\centering
\begin{tabular}{@{}lcccc@{}}
\toprule
                               & VRASED      & SeED & SMARM & Erasmus \\ \midrule
Fresh information              & On-demand   & N/A           & On-demand & ERASMUS+OD* \\   
\makecellspaced{Comprehensive \\ information}     &  HMAC  & HMAC     & HMAC  & HMAC   \\
\makecell[l]{Trustworthy \\ mechanism}          & verified           & X     & X  & X  \\
No leaks                       & secure stack           & X     &  X   & X \\
Exclusive access               & ROM         & X & X & X \\
Immutability                   & ROM         & X & X  & X \\
Atomic execution \vspace{0.1cm}            & by reset    & NMI & X  & X  \\
\makecell[l]{Controlled \\ invocation}          &   Program counter         & X & X & X \\\bottomrule
\end{tabular}
\caption{Hybrid remote attestation protocols}
\label{tab:hybrid_eval}
\end{table}



\begin{description}
    \item[VRASED \cite{vrased}] is the first formally verified \RA protocol~\cite{vrased}. Its attestation mechanism is based on strict program counter logic and the formally verified HMAC algorithm; HACL*. The protocol is verified using Linear Temporal Logic(LTL). The LTL specification describes how interrupts are prevented during execution of the attestation. Therefore, if interrupts are not disabled by software running on \PRV before executing the attestation code, any interrupt that is raised during an attestation will cause a device reset.
    PURE~\cite{pure} extends on VRASED, allowing for secure code update, reset and erasure. The VRASED+PURE \RA architecture has been implemented on low-end, commercially available embedded system such as MSP430.
    
    \item[SeED \cite{seed}] is a non-interactive \RA protocol, meaning that the \PRV has no incoming traffic from the \vrf, which helps mitigate DoS attacks. Instead, the \PRV initiates the attestation protocol based on a Real-Time Clock. SeED is suitable for \RA on resource constrained devices, since it is efficient in terms of power consumption and communication. When the attestation code is triggered a non-maskable interrupt(NMI) is triggered to ensure atomic execution.
    
    \item[SMARM \cite{smarm}] is a light-weight remote attestation protocol to defend against roving malware. That is, malware that moves between locations in memory. Interrupts are allowed, as atomic execution may be too costly for mission critical systems. SMARM uses shuffled measurements as described in section \ref{sec:extensions}. SMARM assumes KFV malware, that only knows the block size of the attestation, and hops to a new memory block after a block of memory has been attested. 
    
    \item[ERASMUS \cite{erasmus}] is an \RA protocol that utilizes periodical self-measurement of \PRV. \VRF then collects and verifies the measurements of \PRV. The primary use case of ERASMUS, is safety-critical systems that do not require on-demand \RA, i.e. fresh information. As seen in table~\ref{tab:hybrid_eval}, we can acquire fresh information with ERASMUS using an extension to ERASMUS: "On-Demand" (ERASMUS+OD). This extension is incorporating an additional attestation at the time of retrieval at the \prv. This may be relevant if real-time \RA is needed, e.g., immediately before or after a software update.
    
\end{description}

Of the four hybrid \RA protocols presented, 
VRASED is the most generic in terms of functionality. VRASED provides general purpose remote attestation with no specific scenario or adversary in mind for low-end embedded computing devices.
Due to the non-interactive nature of SeED, it does not provide on-demand \RA, similarly to ERASMUS. In both protocols, the focus is to accommodate low-end resource constrained devices.
If on-demand \RA is not required, and \PRV is a resource constrained device, SeED and ERASMUS may be suitable.
SMARM is an \RA protocol that utilizes shuffled measurements to defend against roving malware.

\newpage
\subsection{Swarm Attestation}
To evaluate Swarm attestation the evaluation is split into two parts. Initially the requirements set by each protocol are described. Afterwards the functionality that each protocol provides are described. This separation is due to the diversity in requirements and functionality in the space of swarm attestation.

\begin{table}[H]
\centering

\begin{tabular}{@{}lccccc@{}}
\toprule
                               & SEDA & DIAT & SARA & PASTA & PADS \\ \midrule

\makecell[l]{Trust anchor} & X & X & X & X & X \\
Signature scheme & - & X & - & X & - \\
Mac algorithm & X & X & X & X & X \\
Secure Real-time clock & - & - & - & X & X \\
Data-flow monitoring & - & X & - & - & -\\
Control-flow monitoring & - & X & - & - & - \\\bottomrule
\end{tabular}
\caption{Requirements for swarm remote attestation protocols}
\label{tab:swarm_req}
\end{table}

Table \ref{tab:swarm_req} presents the requirements for swarm attestation, the 'X' denotes a dependency. Trust anchor together with the MAC are provided by the underlying security architecture and the underlying implementation of hybrid attestation respectively. The remaining requirements will be described along with the protocols. 
In table \ref{tab:swarm_func} the functionalities of each protocol are described based on properties outlined in section~\ref{sec:properties_of_ra}. 

\begin{table}[H]
\centering
\begin{tabular}{@{}lccccc@{}}
\toprule
                               & SEDA & DIAT & SARA & PASTA & PADS \\ \midrule
Efficiency & Yes & Yes & Yes & Yes & Yes\vspace{0.2cm} \\
\makecellspaced{Aggregation \\ of results} & \makecell{Count of \\ attested devices} & \makecell{Distributed \\ verification} & \makecell{Aggregation at \\ subscribers} & \makecell{Schnorr \\ multisignatures} & \makecell{Swarm \\ consensus} \\ 
\makecellspaced{Attestation \\ topology} & \makecell{Spanning \\ Tree} & \makecell{Spanning \\ Tree} & \makecell{Publish \\ Subscribe} & \makecell{Mesh \\ Network} & \makecell{Any} \\
Fault tolerance \vspace{0.2cm} & - & - & - & Yes & Yes \\
\makecellspaced{Control flow \\ attestation} & - & Yes & - & - & - \\
\bottomrule
\end{tabular}
\caption{Functionality of swarm remote attestation protocols}
\label{tab:swarm_func}
\end{table}

\begin{description}
\item[SEDA \cite{seda}] is among the first \RA protocols that supports scaling networks of devices. SEDA can be built on the same security architectures as other hybrid attestation schemes, and provide the same guarantees. SEDA uses two different protocols during attestation, one from the \vrf to the \prv, and one for the \prv to propagate through the network. This creates a spanning tree that covers the entire swarm, however it is not fault tolerant. The spanning tree of SEDA is built before attestation time and maintains its structure. This in turn means that a node becoming unresponsive will cause all of its children to become unresponsive.

\item[DIAT \cite{diat}] 
is a protocol that, in addition to attesting the software state of a program, also continually monitors the data- and control-flow. Data-flow monitoring ensures that access to information is monitored, while control-flow monitoring ensures that no unprivileged control-flows occur. This ensures that an adversary changing the execution flow but not the program memory is still detected. DIAT uses distributed verification, this means that \emph{Provers} in the swarm also act as \emph{Verifiers}. When an attestation request is received the request is propagated throughout the network in a spanning tree and aggregated as each device attests itself. DIAT uses digital signatures to ensure integrity of attestation results.

\item[SARA \cite{sara}]
is an asynchronous attestation protocol. This is realized through a publish subscribe architecture for attestation. In this protocol the \vrf sends a challenge to a \prv that acts as a publisher in the publish/subscribe architecture. The \prv self attests and publishes the result to subscribers. The subscribers then self attest, combine the results of the attestations and store the result. At a later time the \vrf can query the subscriber to obtain attestation proof for that unit and all publishers it subscribes to.

\item[PASTA \cite{pasta}] is an attestation protocol for swarms where every \prv also acts as a \vrf. PASTA is designed to be deployed in dynamic and changing networks, thus the property of distributed verification provides the network with fault tolerance. The network relies on the Schnorr multi-signature scheme to provide smaller signatures which are aggregated to validate the internal state of the network. PASTA uses distributed verification to allow the network to attest autonomously.
Loosely synchronized real-time clocks are used to ensure that the swarm executes an attestation at certain intervals. Each device measures the time since the last attestation and, if a certain amount of time has passed, initiates an attestation. The attestation can be initiated at any \prv and is propagated through the swarm using a mesh network of devices that is either created at swarm initialization or is updated dynamically. If the mesh network allows for dynamic updates, PASTA has fault tolerance. PASTA with fault tolerance allows devices to leave the network between attestations.
There is one drawback to the PASTA protocol, if a device is absent from the network for longer than the time between attestations it may be regarded as untrustworthy.
If an external \vrf wants to check the state of the network and evaluate the attestations, the \vrf needs to reconnect at certain intervals depending on the allowed time for a device to be offline. There are propositions to fix this; either to rely on existing techniques to detect hardware attacks, or extending the protocol to provide physical tamper evidence in which case a device can be determined to be benign.

\item[PADS \cite{pads}]
is a protocol based on the principle of non-interactive attestation, similarly to SeED (section \ref{eval_hybrid}). However instead of communicating the result to a \vrf, the device attests itself using a pre-programmed set of valid states at a certain time $t$. All devices in the swarm attest simultaneously due to the assumption of a secure real-time clock. The swarm then uses a minimal consensus algorithm to distribute a snapshot of the swarm state to all devices. A \vrf can then query any device in the swarm to get a snapshot of the swarm.
PADS can be implemented with any topology of the attestation including highly unstructured networks. This is due to the consensus based aggregation of the attestation.
    
\end{description}

%

Each of the evaluated protocols for attesting collections of devices solve a different issue. SEDA was the first protocol for this purpose specifically, and has very few requirements, however it does provide strong guarantees for stable swarms. DIAT is a framework that provides many features such as control-flow monitoring to validate the run-time state, and not just the memory. SARA provides an asynchronous attestation protocol using a publish subscribe system which is a method of aggregation distinct from the rest. Pasta uses multi-signatures to aggregate attestation results, and PADS uses a consensus algorithm. Each of these serve a distinct use case and have their own merit.


\newpage
\section{Conclusion}\label{sec:conclusion}
To present a holistic view of state-of-the-art in remote attestation, we have conducted a literature review.
We have uncovered three overall paradigms of remote attestation; software-based, hardware-based, and hybrid. Furthermore, we looked at which adversary models are used when describing remote attestation. Lastly, we have investigated extensions such as remote code updates and control-flow attestation that allows a \vrf to reason about the run-time state of a \prv. 

Software-based attestation provide the possibility of attesting a device without depending on any hardware modules on the device. The guarantees of software based attestation is dependent on strong assumptions about adversary capabilities. However, software-based attestation is a cheap and easy to implement form of remote attestation that provide minimal guarantees in one-hop communication distance. 
Hardware-based remote attestation provides strong guarantees by using secure hardware. However it requires specialized hardware that is expensive to deploy on many devices and thus see a more limited set of use-cases.
Hybrid attestation bridge the gap between software and hardware by leveraging security architectures to achieve the necessary guarantees to provide secure remote attestation. This allows hybrid attestation to remain inexpensive and secure.

Swarm attestation is independent of the underlying paradigm of remote attestation. It allows a \vrf to reason the internal state of a collection of \prv devices in a way that is more efficient than attesting each device individually. In swarm settings the entirety of a swarm can be verified, however the state of the individual devices cannot be reasoned about due to result aggregation. 

Furthermore, TOCTTOU attacks are commonly considered in \ra. It is an attack where an adversary will save a valid snapshot of a device state, and present it as a current state in a later attestation request. 

Some of the latest additions to the field of remote attestation include formal verification of remote attestation protocols where VRASED is the only formally verified protocol so far. 
Dynamic swarm networks which allow devices to leave and join networks, as well as distributed verification in swarm settings. Some of the problems that remain in remote attestation are (1) identification of faulty devices in swarm attestation settings (2) Further development of dynamic swarm topologies (3) Further formal verification of protocols to provide provably secure remote attestation protocols.

\newpage
\bibliographystyle{unsrt}
\bibliography{main}


\end{document}